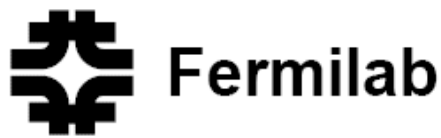

# CODE INTERCOMPARISON AND BENCHMARK FOR MUON FLUENCE AND ABSORBED DOSE INDUCED BY AN 18-GEV ELECTRON BEAM AFTER MASSIVE IRON SHIELDING[*†]

Alberto Fasso[1], Alfredo Ferrari[2], Anna Ferrari[3], Nikolai V. Mokhov[4], Stefan E. Mueller[#3], Walter Ralph Nelson[5], Stefan Roesler[2], Toshiya Sanami[6], Sergei I. Striganov[4], Roberto Versaci[1]

[1]ELI Beamlines, Czech Republic
[2]CERN, Switzerland
[3]Institute of Radiation Physiscs, Helmholtz-Zentrum Dresden-Rossendorf, Germany
[4]Fermilab, Batavia, IL 60510, U.S.A.
[5]SLAC National Accelerator Laboratory (retired), USA
[6]KEK, Japan

## Abstract

In 1974, Nelson, Kase, and Svenson published an experimental investigation on muon shielding using the SLAC high energy LINAC. They measured muon fluence and absorbed dose induced by a 18 GeV electron beam hitting a copper/water beamdump and attenuated in a thick steel shielding. In their paper, they compared the results with the theoretical models available at the time. In order to compare their experimental results with present model calculations, we use the modern transport Monte Carlo codes MARS15, FLUKA2011 and GEANT4 to model the experimental setup and run simulations. The results will then be compared between the codes, and with the SLAC data.

[*]Work supported by Fermi Research Alliance, LLC under contract No. DE-AC02-07CH11359 with the U.S. Department of Energy.
[†]Presented paper at the 12th Workshop on Shielding Aspects of Accelerators, Targets and Irradiation Facilities, SATIF-12, Fermilab, April 28-30, 2014.
[#]stefan.mueller@hzdr.de

# Code intercomparison and benchmark for muon fluence and absorbed dose induced by an 18 GeV electron beam after massive iron shielding

Alberto Fassò[1], Alfredo Ferrari[2], Anna Ferrari[3], Nikolai V. Mokhov[4], Stefan E. Müller[3], Walter Ralph Nelson[5], Stefan Roesler[2], Toshiya Sanami[6], Sergei I. Striganov[4], Roberto Versaci[1]

[1]ELI Beamlines, Czech Republic
[2]CERN, Switzerland
[3]Institute of Radiation Physiscs, Helmholtz-Zentrum Dresden-Rossendorf, Germany
[4]Fermi National Accelerator Laboratory, USA
[5]SLAC National Accelerator Laboratory (retired), USA
[6]KEK, Japan

**Abstract**

*In 1974, Nelson, Kase, and Svenson published an experimental investigation on muon shielding using the SLAC high energy LINAC [1]. They measured muon fluence and absorbed dose induced by a 18 GeV electron beam hitting a copper/water beamdump and attenuated in a thick steel shielding. In their paper, they compared the results with the theoretical models available at the time. In order to compare their experimental results with present model calculations, we use the modern transport Monte Carlo codes MARS15, FLUKA2011 and GEANT4 to model the experimental setup and run simulations. The results will then be compared between the codes, and with the SLAC data.*

**Introduction**

It has been stressed recently [2] that a good understanding of muon photoproduction by high-energy, high-intensity beams of photons or electrons is crucial in the shielding design for future beam facilities like the ELI beamlines facility in Prague [3]. The same holds for upcoming projects at LCLS (SLAC) and the planned ILC in Japan. The underlying question is how well the theoretical models for photoproduction of muons in current transport codes used for shielding design represent experimental data. We have therefore started to prepare benchmark calculations using the codes MARS15, FLUKA2011 and GEANT4 in order to compare them to data from an experiment done in 1974 at the Stanford Linear Accelerator Center in California. This will also lead to a code intercomparison concerning the implementation of muon photoproduction models in the different codes.

*The experiment*

In 1974, Nelson, Kase, and Svenson carried out an experimental investigation at SLAC to study the muon fluence and absorbed dose induced by an 18 GeV electron beam hitting a copper/water beamdump [1]. In the vicinity of a nucleus, the electrons produced bremsstrahlung photons in the beamdump, which subsequently lead to muon pair photoproduction. The muons were produced within 6 radiation lengths in the beamdump (corresponding to 22.23 cm), and were subsequently attenuated by thick blocks of shielding iron. The lateral distribution of the muon fluence and the absorbed dose were measured by positioning detectors perpendicular to the incident electron beam axis in four narrow gaps (gap A, gap B, gap C, gap D) between the iron shielding blocks. The muon fluence was detected using 400μm thick nuclear track emulsion plates, which were read out by microscopes after the exposure. Thermoluminescent dosimeters were used to register the absorbed dose. In addition, two scintillation counters determined the exposure and also cross-checked the muon fluence measurements. The geometry of the setup allowed to perform measurements at vertical angles from 0 to 150 milliradians. Within this range, it is ensured that the direct flight paths from the muon production point in the beam dump to their detection are completely contained in the iron shielding. To protect the detectors against background radiation, the gaps A, B and C were covered with lead blocks on the side and on top. Gap D, which is the furthest away from the muon production point, was left exposed.

The authors compared their experimental results to the theoretical formulation derived in [4]. It was found that the muon fluence and absorbed dose were accurately predicted by the theory for polar angles below 30 milliradians, while for larger angles, the theoretical prediction was underestimating both fluence and dose by an order of magnitude or more.

**Transport codes**

In 2007, a first comparison [5] was undertaken of the experimental data with the modern transport codes MARS [6-10] and FLUKA [11][12]. Together with the observations made in the shielding design for the ELI beamlines facility in Prague [3], this inspired the idea to redo the comparison with the newest versions of the two codes, and also include the GEANT4 [13][14] toolkit as a third transport code into the comparison.

*MARS15*
The MARS code is a general-purpose, all-particle Monte Carlo simulation code. It contains established theoretical models for strong, weak and electromagnetic interactions of hadrons, heavy ions and leptons. Most processes can be treated either exclusively (analogously), inclusively (with corresponding statistical weights) or in a mixed mode. There are several options for the geometry, with "extended"or ROOT-based [15] modes as the most commonly used ones. Photoproduced muons are included in the MARS code in two ways:

- An exclusive muon generator based on the Weizsäcker-Williams approximation using algorithms based on the work of [16]. Only coherent photomuon production is simulated. This generator is used as the default generator for muon production.

- An inclusive muon generator based on the calculation of the lowest-order Born approximation in [17][18] for targets of arbitrary mass, spin and form factor as well as arbitrary final states.

Both models give practically identical results for photon energies larger than 10 GeV. At lower energies, a precise description of the nuclear form factors becomes important. MARS supports two options for the description of the nuclear density for the inclusive muon generator: the original Tsai power-law mode and a symmetrized Fermi function. Angular and momentum distributions of muons produced by bremsstrahlung photons of 18 GeV electrons in copper simulated with the inclusive and the exclusive generator are in close agreement. The Weizsäcker-Williams approximation is therefore adequate for the benchmark in question.

*FLUKA2011*
FLUKA is a fully integrated particle physics Monte Carlo simulation package containing implementations of sound and modern physical models. A powerful graphical interface (FLAIR [19]) facilitates the editing of FLUKA input, execution of the code and visualization of the output. Photomuon production in FLUKA is implemented in the Weizsäcker-Williams approximation using the formalism of [17][18]. Only coherent photomuon production is simulated.

*GEANT4.10*
The GEANT4 toolkit is the successor of the series of GEANT programs for geometry and tracking developed at CERN. It is based on object-oriented software technology. GEANT4 represents a set of software tools from which the user needs to program his own application. The implementation of muon pair production is described in [16] and is based on the work in [20].

The geometry of the experiment has been modelled using the three codes based on information from [1][21]. Vertical views of the resulting geometries can be seen in Figure 1 (MARS), Figure 2 (FLUKA) and Figure 3 (GEANT4). Figure 1 also indicates the location of the beamdump and the gaps in which the detectors were placed in the experiment. The electron beam is coming from the left and hit the beamdump. While the geometries in Figures 1 – 3 still show some minor differences concerning dimensions and material between the different codes, a consistent geometry was defined at the SATIF-12 workshop, which will be used in upcoming simulation campaigns with all three codes.

**Scoring and simulation parameters**

In order to score the results with the different transport codes and compare with the experimental results, the following scorers were defined:

- The **muon fluence** in the 4 gaps normalized to the integrated electron charge on the beam dump (in $\mu/cm^2/$Coulomb)

- The **absorbed dose** in the 4 gaps normalized to the integrated electron charge on the beam dump (in rad/Coulomb). To simulate the dose deposition in the thermoluminescent dosimeters, thin layers of LiF (500µm thickness) are placed in each gap.

- Several double-differential scorers in **energy and angle** for muons crossing the copper-water intersections over approximately 6 radiation lengths in the beamdump allow to cross check the implementation of muon photoproduction in the different codes.

For MARS, scoring distributions are obtained via post-processing using PAW. FLUKA scoring distributions are obtained from the built-in scorers and post-processing is done using FLAIR [20]. With GEANT4, a mixture of built-in scoring and ROOT histograms [15] is used.

The following simulation parameters and configuration options are used in the simulations:

1. MARS (used version: MARS15 (2014)):

    – Generation and transport thresholds were set at 2 GeV in the beamdump, at 1 GeV in the shielding upstream, and at the following values elsewhere: 0.001 eV for neutrons, 1 MeV for muons, heavy ions and charged hadrons, 0.1 MeV for photons, and 0.2 MeV for electrons

2. FLUKA (used version: FLUKA2011.2b.6):

    – Defaults for precision simulations are used

    – Production and transport thresholds for electrons and photons are set to 100 keV and 10 keV, respectively

    – Full simulation of muon nuclear interactions and production of secondary hadrons switched on

    – Production of secondaries for muons and charged hadrons switched on (100 keV threshold)

3. GEANT4 (used version: GEANT4.10p1):

    – Basic physics list with quark gluon string and Bertini models is used, with parameters for electromagnetic physics tuned for high precision

    – Additional process for gamma conversion to muons switched on for photons

    – Additional process for muon-nucleus interactions switched on for muons

    – Range threshold for gamma, electron, positron, proton: 700µm

**Preliminary results**

Up to now, preliminary results exist only for FLUKA-based simulations. Simulation campaigns with MARS and GEANT4 are in progress, but no results are available yet. Figure 4 shows the muon fluence registered in the four gaps for data (black triangles) compared to the FLUKA simulation (red squares) with 5 million simulated events. The agreement between the simulated and experimental data sets is quite good. The same holds for Figure 5, in which the absorbed dose is plotted for data and FLUKA simulation (same symbols and colorcodes as in Figure 4 apply), except for gap D, where the lack of lead shielding in the experiment leads to a constant offset in the data for angles above 60 milliradians, which was already noted in [1]. In particular, it can be concluded that both in Figure 4 and Figure 5, an underestimation of both fluence and dose for angles above 30 milliradians, as reported in [1], is not observed. The simulations in Figure 4 and Figure 5 were done using cast iron with a density $\rho=7.0$ g/cm$^3$ for the material of the shielding blocks. Due to the fact that the exact density of the material used to construct the shielding blocks is not known at the moment, additional simulations were done using steel with a density of $\rho=7.6$ g/cm$^3$ as shielding material. The corresponding results for gap A are shown in Figure 6 (muon fluence) and Figure 7 (absorbed dose). A clear effect of the different materials is visible only for muon fluence at angles < 60 milliradians, with steel giving a lower value for muon fluence than iron. For larger angles and for the absorbed dose, the effect is masked by the current statistical uncertainties of the simulations.

**Summary and conclusions**

Based on the experimental results on muon production by an 18 GeV e$^-$ beam hitting a copper-water target reported by Nelson, Kase, and Svensson, the Monte Carlo transport codes MARS, FLUKA and GEANT4 have been used to model the experimental conditions. First preliminary results on muon fluence and absorbed dose have been produced with FLUKA. The agreement between the simulated results and the experimental values is quite promising. Results with the Monte Carlo transport codes MARS and GEANT4 are expected to follow soon. Some uncertainties about actual geometry and material composition still exist. The original logbooks of the experiment, which are currently being retrieved, may be of help here. Further refinement of the simulations together with consistency checks will allow to compare the implementation of muon production and transport in the different Monte Carlo codes.

**Figure 1: Geometrical model of the experimental setup using MARS15**

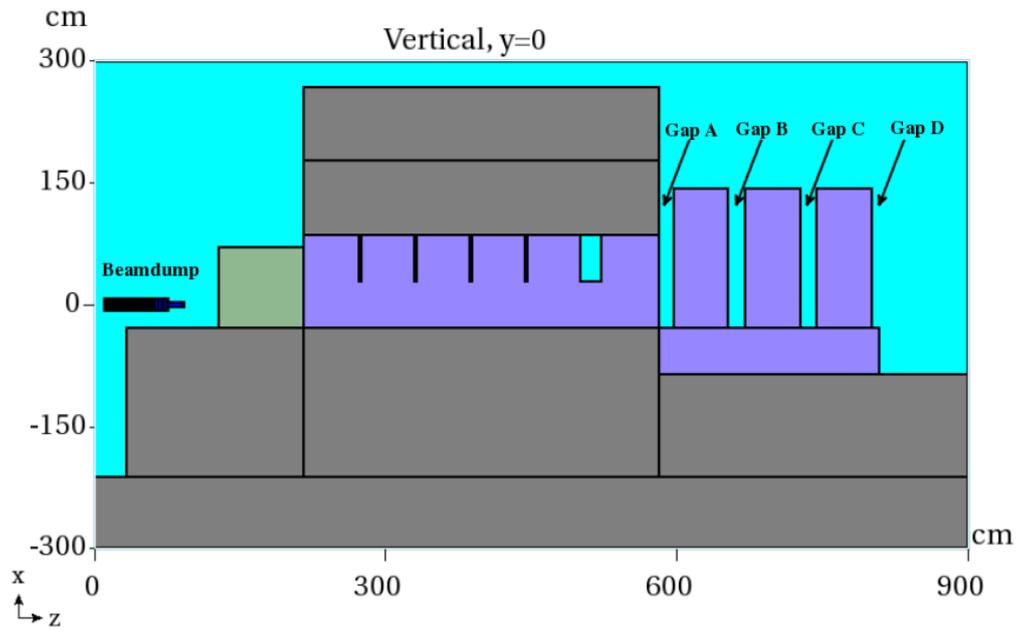

**Figure 2: Geometrical model of the experimental setup using FLUKA2011**

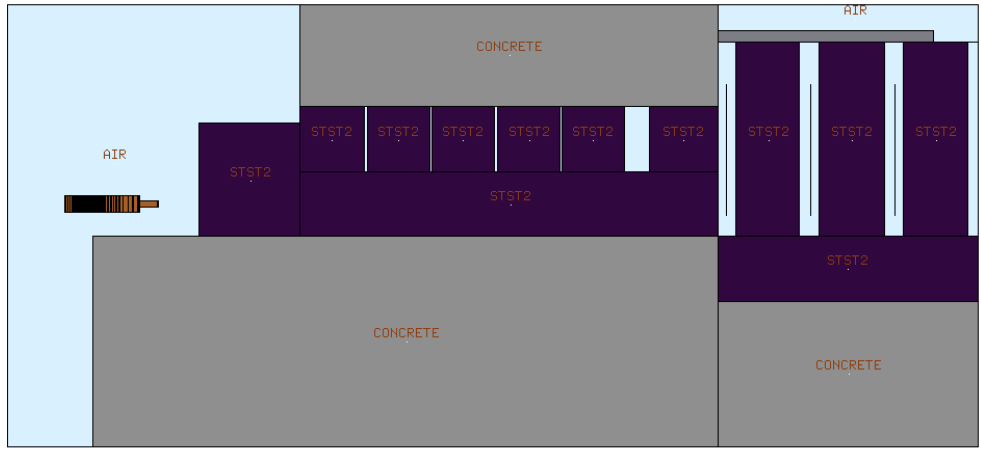

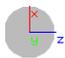

**Figure 3: Geometrical model of the experimental setup using GEANT4.10**

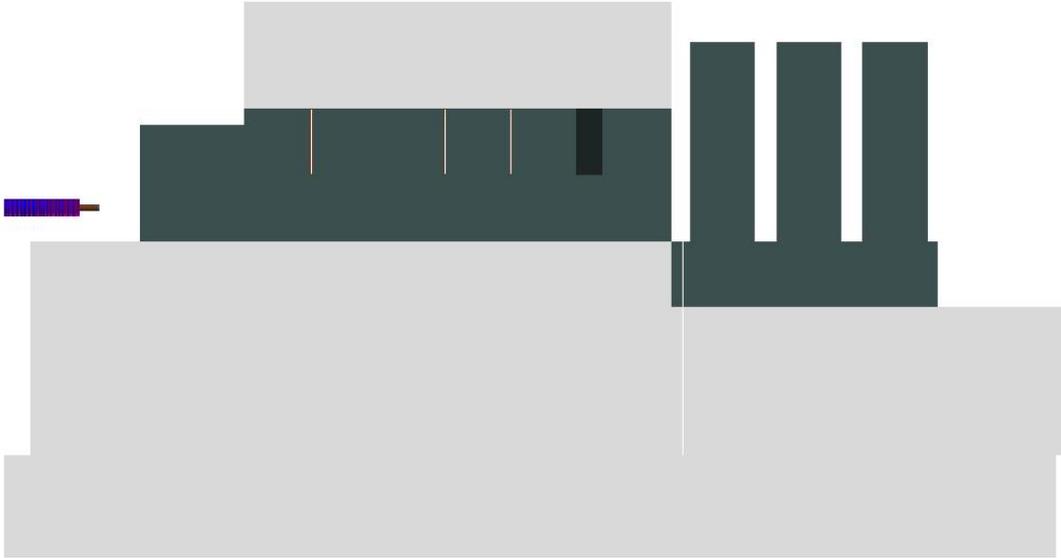

**Figure 4: Preliminary results for muon fluence in gap A-D for FLUKA simulation compared to data**

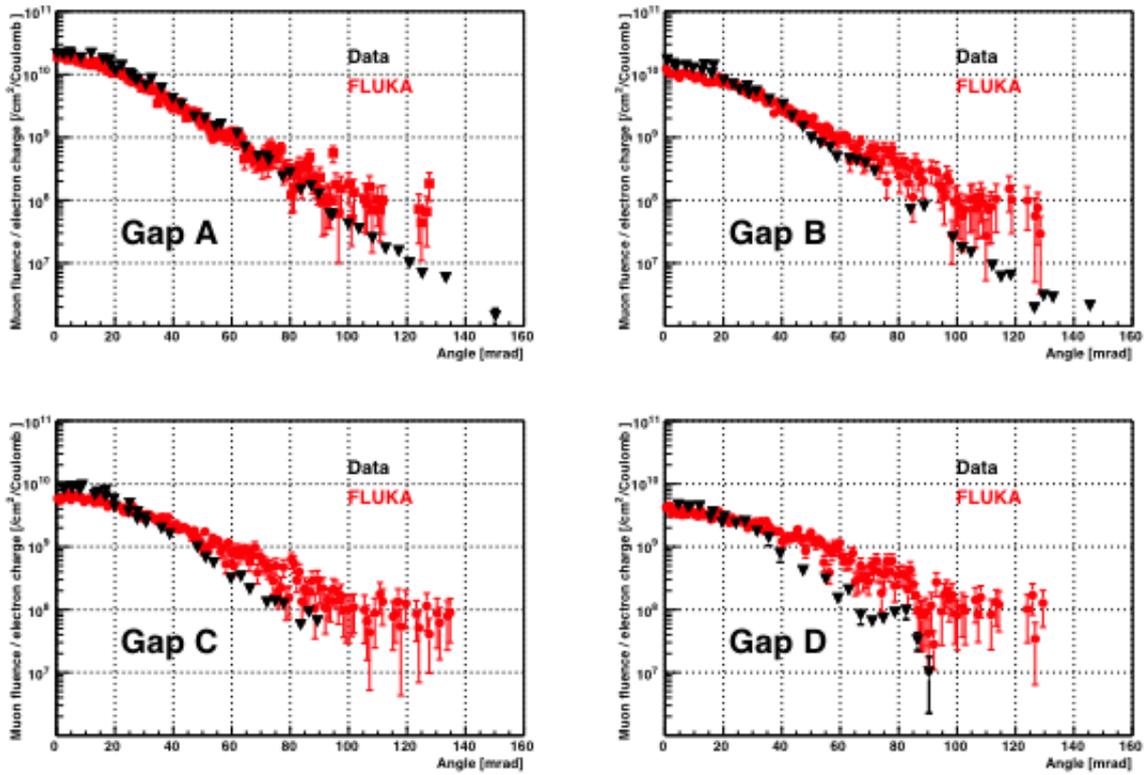

**Figure 5: Preliminary results for absorbed dose in gap A-D for FLUKA simulation compared to data**

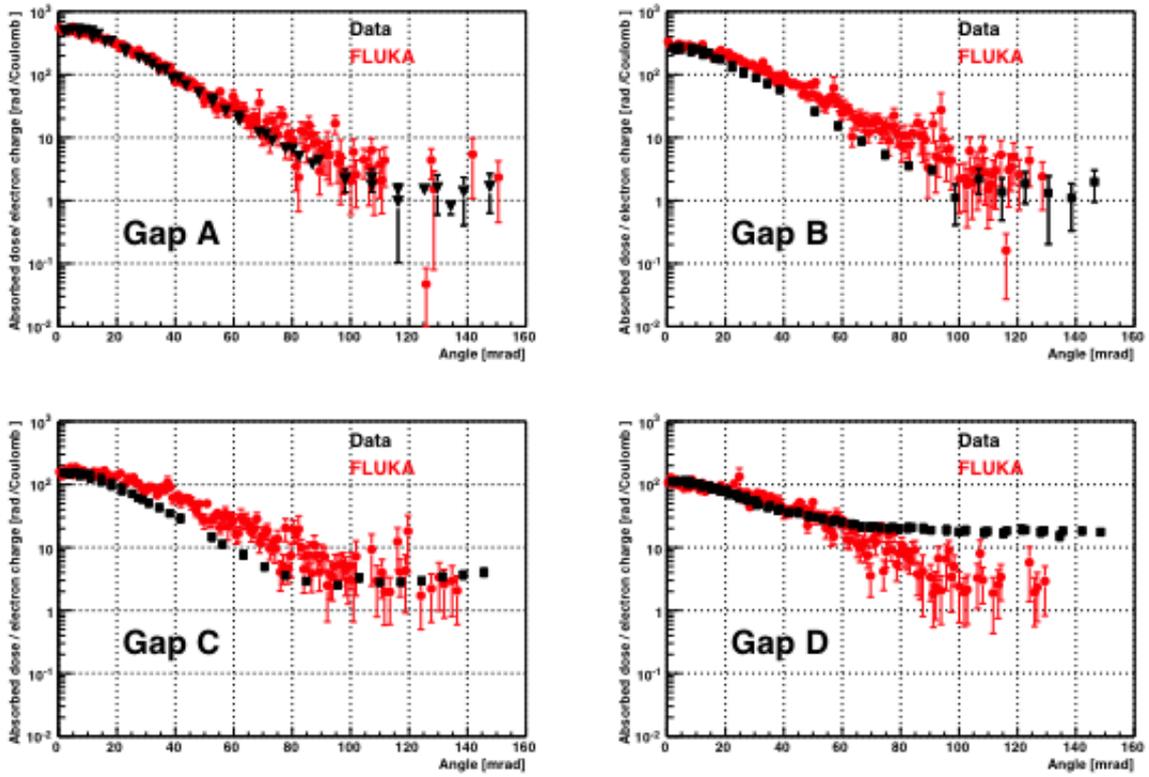

**Figure 6: Preliminary results for muon fluence in gap A for FLUKA simulations and data**

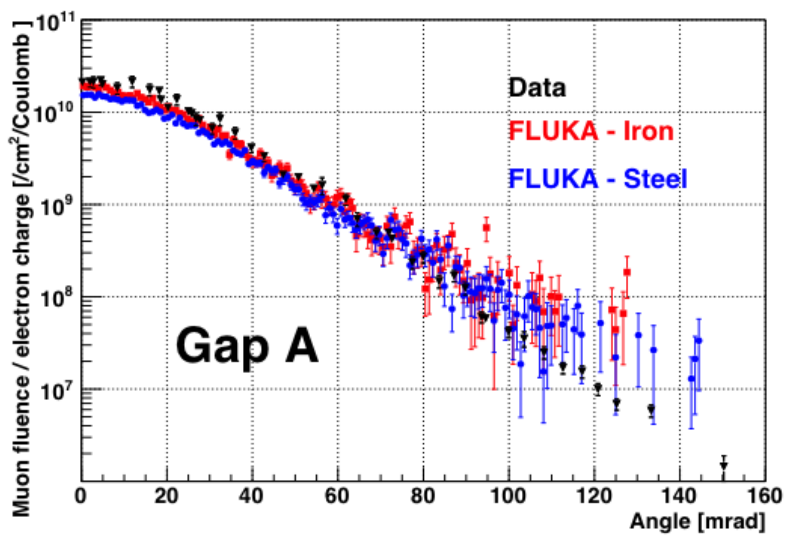

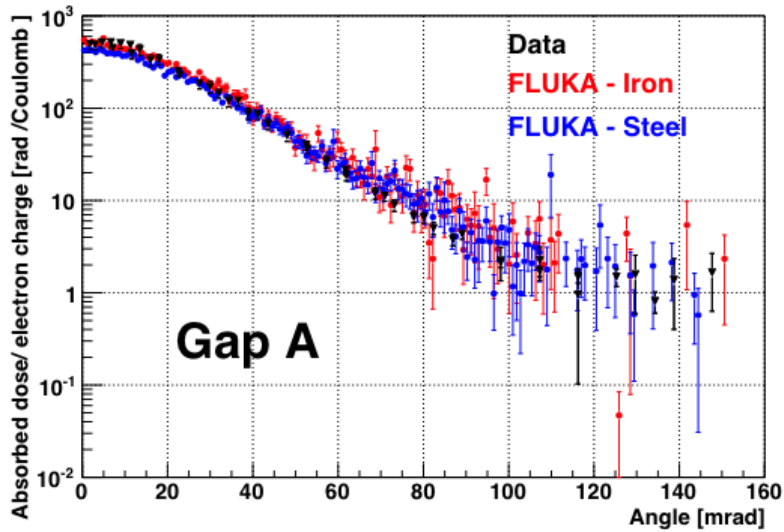

**Figure 7: Preliminary results for absorbed dose in gap A for FLUKA simulations compared to data**